\documentclass[prl,twocolumn]{revtex4-2} 

\newcommand{\EF}{{\cal E}}

\usepackage{graphicx}
\usepackage{amssymb}

\newcommand{\ket}[1]{ \left| #1 \right> }
\newcommand{\bra}[1]{ \left< #1 \right| }
\newcommand{\braket}[2]{ \left< #1 | #2 \right> }

\newcommand{\cf}[1]{\braket{#1}{\rm vac}}

\newcommand{\be}{\begin{eqnarray}}
\newcommand{\ee}{\end{eqnarray}}

\usepackage{pgfplots}
\usetikzlibrary{pgfplots.groupplots}
\usetikzlibrary{shapes,arrows}
\pgfplotsset{compat=1.18}
\usepackage{tikz}

\newcommand{\myfig}[2]
{ \centerline{\resizebox{!}{#1\textwidth}{\includegraphics{#2}}} }

\begin{document}

\title{Unlocking vacuum entanglement}

\author{Andrew Steane and Haru Ishizaka}

%\email{a.steane@physics.ox.ac.uk} % optional

\affiliation{Department of Atomic and Laser Physics, Clarendon Laboratory, Parks Road, Oxford OX1 3PU, England}

\date{\today}
%\label{firstpage}

\begin{abstract}
The structure of entanglement in the ground state of the harmonic chain is studied. A class of two-mode squeezed states, useful for this purpose, is identified.
The entanglement of the local modes at the ends of the chain, after tracing out the centre, rapidly falls to zero as the length of the chain increases. However, if the central modes are measured, and the result communicated to systems interacting with the outer modes, the latter exhibit greatly enhanced entanglement, including in conditions where none was otherwise available.
These ideas can be demonstrated in experiments in trapped ions, among other systems. The  extension to the continuous case yields enhanced entanglement extracted from the vacuum state of a bosonic quantum field.
\end{abstract}
\maketitle

There is now a large literature on vacuum entanglement \cite{Unruh1976,Summers1987,Reznik2003,Reznik2005,Gao2025}.
In this paper we consider
entanglement properties of the ground state of 
the harmonic chain (a chain of interacting harmonic oscillators)
\cite{Audenaert2002,Botero2003,Botero2004}. The
results extend also to a string of ions in a linear trap, and to the vacuum state of a bosonic quantum field theory. We focus on entanglement between degrees
of freedom at the two ends of the chain, especially the fact that this can
be greatly enhanced by classical communication of the results of observations on the central modes.

\begin{figure}
	\myfig{0.25}{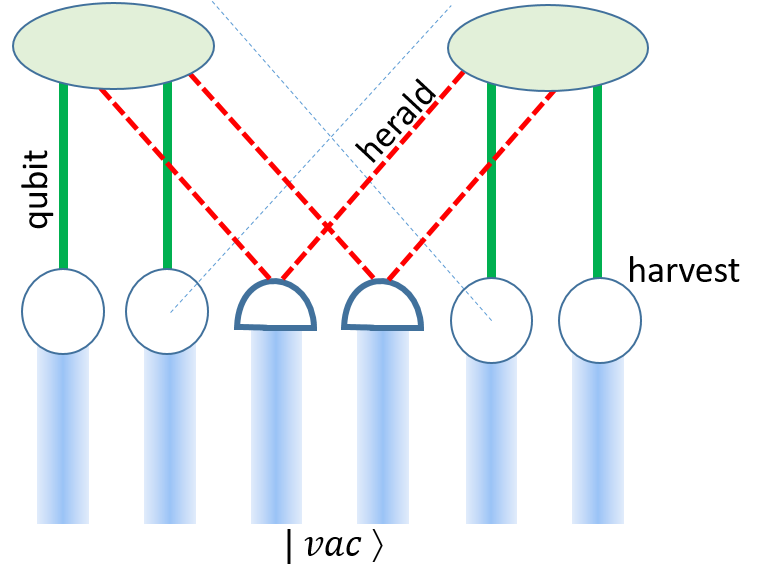}
	\caption{Spacetime diagram showing the overall scenario studied in the text.
		A $N$-mode harmonic chain ($N=6$ is illustrated) is in its ground state.
		The outer modes are harvested into 2 or more qubits (green lines). The
		central modes are measured in the number basis and the results communicated
		to the qubits. The latter finish in an entangled state.}
	\label{f.setup}
\end{figure}

The situation is comparable to the following well-known case. If three parties 
$A,B,C$ share the GHZ state $(\ket{000} + \ket{111})/\sqrt{2}$ then the 
entanglement between any pair of the parties, in the absence of information from 
the third, is zero (for, a projective measurement of any one of the qubits in 
the computational basis will leave the others in a product state). 
If, on the other hand, the central party $B$ measures their qubit in the basis
 $(\ket{0}\pm \ket{1})/\sqrt{2}$ then the remaining qubits are projected onto either 
 	$(\ket{00}+\ket{11})/\sqrt{2}$ or $(\ket{00}-\ket{11})/\sqrt{2}$, with equal probability. There is
still no entanglement between $A$ and $C$ in the absence of any communication from $B$. But
if $B$ now communicates to $A$,$C$ the outcome of his observation, then $A$,$C$ learn which
state they have and now they have 1 ebit of entanglement between them. This scenario is somewhat like,
but different from, distillation of entanglement. It is a kind of activation or unlocking of
entanglement. We will show that a scenario like the above arises in the ground state of the
harmonic chain, and, by extension, in other qualitatively similar systems. By observing the
central modes of the chain, and communicating the results, one unlocks or makes available entanglement
between the outer modes of the chain (c.f. Fig. \ref{f.setup}).

In the physics of quantum information, when a desired outcome is obtained
stochastically, and there is a signal that indicates the occasions when it
has occurred, it is said to be {\em heralded}. 

{\em Entanglement measures}.
We quantify entanglement by calculating the {\em entanglement of formation} $\EF$. 
For a pure state of a pair of systems of any size, this is given by the
{\em  entanglement entropy} (also called {\em von Neumann entanglement} 
or {\em entropy of entanglement}), which is the von Neumann entropy of one system after tracing over the other.
For a mixed state of a pair of qubits we employ the Wootters formula \cite{Wootters1998}. For larger systems
we obtain upper and lower bounds on $\EF$ by numerical search.
$\EF$ is fungible;
for any given value one may meaningfully speak of ``$\EF$ ebits" \cite{BkNielsen}.
We also obtain a quantity which we shall call the {\em harvested entanglement}. 
This is the entanglement of formation of two or more qubits which have interacted with the system in question through some local {\em harvesting
 operation} \cite{Martin-Martinez2016,Gooding2024,Lindel2024}. The harvested entanglement $\EF_v$ is obtained from the qubits' reduced density matrix after tracing out the system. 
We adopt as harvesting operation
$
U_v =  \exp\left[i(\pi/2)(\sigma a^\dagger +\sigma^\dagger a)\right]
$
where $\sigma = \ket{g}\bra{e}$
and $a$ is the lowering operator for the {\em local mode} of the atom (in a harmonic chain) or ion (in a trap) (the local modes are described below).
For a qubit prepared in $\ket{g}$ and motional number states $n < 2$, $U_v$ enacts a `swap' between the motion and qubit. A method to produce this operation
rapidly 
by laser pulses on an atom with suitable electronic structure is described
in \cite{Retzker2005}. Ideally the operation should be completed on a timescale
fast compared to the motional frequency, so that any entanglement arriving
in the qubits cannot be attributed to signalling down the chain at the speed of sound. We shall assume that in the continuous limit of a chain---the
quantum field---a detector can be employed whose interaction with the local field is similar to this after averaging over a set of adjacent modes.
Different detector interactions are spacelike-separated in the case of 
the electromagnetic field, and separated by $\Delta x > v _s\Delta t$
more generally, with $v_s$ the speed of signals in the field.

We are concerned with a case where a given bipartite system may be projected by 
measurement onto one of a set of bipartite pure states, with probabilities $p_i$,
and the measurement outcome is known. If the $i$'th such
state has entanglement $\EF_i$ then the entanglement available to be used,
given that the measured outcome was $i$, is just $\EF_i$. 
The expected value (or average) of the entanglement obtained this way is
$\bar{E} = \sum_i p_i \EF_i$. We shall call this the {\em heralded} entanglement; it is a major
focus of this paper.

We now introduce a class of two-mode state not previously studied (to our
knowledge), which we name the {\em two-mode squeezed state of order $k$}, defined as
\be
\ket{\sigma_k(\beta,\theta)} 
&=& \sqrt{1 - e^{-2\beta}} \sum_{n=0}^\infty e^{-n\beta}
\left[ \cos(\theta) \ket{n+k}\ket{n} \right. \nonumber\\
&& \left. + \sin(\theta) \ket{n}\ket{n+k} \right]
\ee
In the case $k=0$ this reduces to the widely studied two-mode squeezed state 
(TMSS), and for that case one should take $\theta=0$ so that $\ket{\sigma_0}$ is 
normalized. For $\theta$ equal to a multiple of $\pi/2$ (and any $k$) the 
entanglement entropy of $\ket{\sigma_k(\beta,\theta)}$ is 
${\EF} = (\lambda+1/2)\log_2(\lambda+1/2)
-  (\lambda-1/2)\log_2(\lambda-1/2)$ where $\lambda = 1/(2 \tanh\beta)$.
For the {\em balanced} case, $\theta = \pi/4$ and $k>0$, $\EF$ is {\em lower}-bounded by 1. Whereas the entanglement of the ordinary TMSS tends to zero at large $\beta$ (low `temperature'), the entanglement of the balanced TMSS of order $k>0$ tends to 1 at large $\beta$. More generally, for $k > 0$ 
the entanglement varies approximately sinusoidally, as function of $\theta$, between the $k=0$ value and a value exceeding 1 (falling near to 1 when $\beta \gtrapprox 2$).

\begin{figure}
\begin{tabular}{cc}
\resizebox{!}{0.25\textwidth}{\includegraphics[trim=3cm 1mm 3cm 1mm, clip=true]{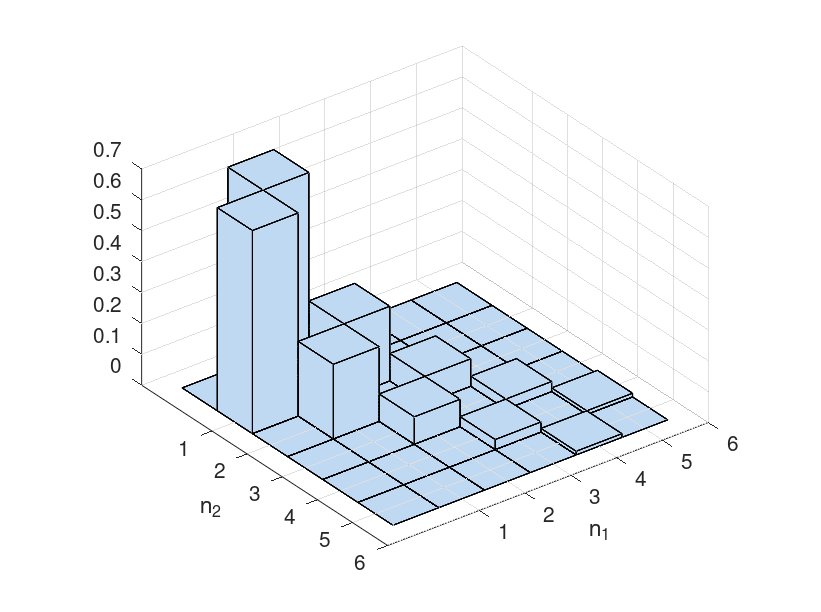}}  
&
\resizebox{!}{0.2\textwidth}{\includegraphics{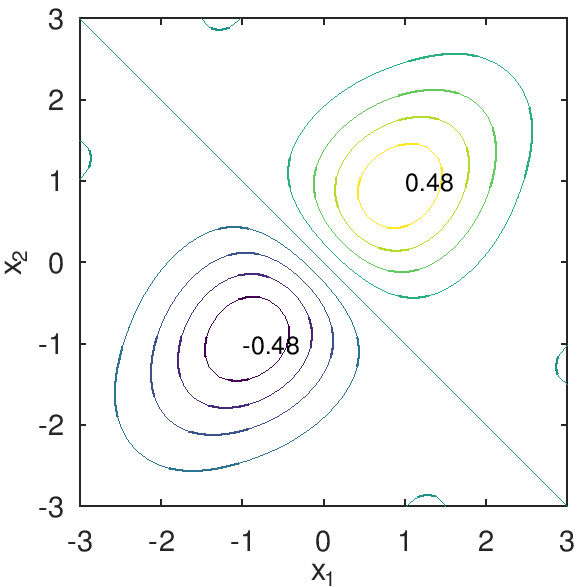}}  % The .png file was generated by show_twomode.m
\end{tabular}
\caption{Balanced two-mode squeezed state of order 1. (a) shows the
coefficients $\braket{n_1,n_2}{\sigma_1(\beta,\pi/4)}$ arranged as a table,
for $\beta = 1$.
(b) Shows contours of the wavefunction in position space, with $x_1$ and $x_2$ the excursions for the two oscillators.}
\label{f.tmss}
\end{figure}

The balanced TMSS of order 1 is visualized in Fig.~\ref{f.tmss}.

We now discuss the harmonic chain of $N$ equal masses and springs. 
We discuss the case $N=3$ in detail, and then proceed to higher $N$.

For any $N$ the ground state of the system has the wavefunction
\be
\psi(q_1, q_2, \cdots q_N) = \prod_{j=1}^N (\nu_j/\pi)^{1/4}  e^{- q_j^2 \nu_j / 2} \label{ground}
\ee
where $q_j$ are the normal mode coordinates, $\nu_j$ the normal mode
frequencies, and we have taken $\hbar = m = 1$ where $m$ is the mass of
one oscillator. One can express this same state in terms of any 
basis. A convenient basis is furnished by a set of {\em local modes},
such that each local mode concerns the state of motion of just one of
the oscillators and furnishes a complete set of states for that oscillator.
The frequencies $\omega_k$ of these local modes can be chosen as we like:
no matter what choice is made, the associated number states furnish a complete 
orthonormal set. A judicious choice will simplify the analysis and
maximise the entanglement to be described. 

For the case $N=2$ a convenient choice of local mode frequencies is 
$\omega_1 = \omega_2 = \sqrt{\nu_1 \nu_2}$. With this choice, the ground
state of the system takes the form of a TMSS when expressed in local
modes \cite{Milburn1984,Caves1985,Caves1991,Retzker2005,Tserkis2017}. With $\nu_2 = \sqrt{3} \nu_1$ (which happens in the
chain and in the ion trap) one finds $\lambda = 0.51898$ and
$\EF = 0.1362$ e-bit. The harvested entanglement is then ${\EF}_v = 0.1315$.

We can understand the choice $\omega_1 = \omega_2 = \sqrt{\nu_1 \nu_2}$
(at $N=2$) by noting that these values result in 
$\cf{2,0} = \cf{3,1} = 0$ (as well as various other zeros)
where the notation gives $\bra{n_1,\, n_2}$ for
local-mode number states and $\ket{\rm vac}$ is the ground state of
all the normal modes, (\ref{ground}).

Proceeding now to $N=3$, one finds that the coefficients $\cf{n_1,n_2,n_3}$
are zero whenever $n_1+n_2+n_3$ is odd. The entanglement harvested
from the outer local modes, tracing over the central local mode, is
of order ${\EF}_v = 0.004$, depending on the local mode frequencies. 
For $N > 3$ this number falls to zero; this is the well-known rapid
reduction of vacuum entanglement with distance. 

For $N=3$ it would be technologically
challenging to produce harvesting operations precise enough to recover
the small entanglement of the outer modes. However, on examining the 
entanglement structure in more detail, a striking fact emerges.
We adopt local mode frequencies such that $\cf{200} = \cf{310} = 0$;
for a 3-oscillator chain this occurs when
\be
\omega_1^2 = \omega_3^2 = \frac{ 2\nu_1\nu_2\nu_3 + (\nu_1\nu_2 + \nu_2\nu_3)\omega_2 }
{ \nu_1 + nu_3 + 2 \omega_2 }  .        \label{w1}
\ee
With this choice and $\omega_2 = 0.86\, \nu_1$ we find
\be
\braket{n_2=0}{\rm vac} &\simeq&
0.956 \ket{\sigma_0(\beta_0)}_{1,3} ,   \label{sig0} \\
\braket{n_2=1}{\rm vac} &\simeq&
0.199 \ket{\sigma_1(\beta_1,\pi/4)}_{1,3}   \label{sig1} 
\ee
where subscripts on kets indicate the local oscillators (e.g. single atoms;
the central atom is number 2) and 
$\{\beta_0,\;\beta_1\} \simeq \{3,\;2.72\}$. 
Consider the state of the outer modes when the central mode
is projected onto a number state. For central $n_2=0$, the
outer modes are in $\ket{\sigma_0(\beta_0)}$, and for
$n_2=1$ they are in $\ket{\sigma_1(\beta_1,\pi/4)}$.
These states have 
${\EF}=0.0251,\; 1.0003$, respectively, and
${\EF}_v = 0.0250, 0.993$.
It follows that if a central observer were to measure the central mode 
in the number 
basis and inform two outer observers of the result, the latter would be able
to harvest much more entanglement than the maximum $\sim 0.0044$ available
without this information. With probability $0.91$ they can obtain an
increase by a factor $0.025/0.0044 = 5.7$, and with probability $0.04$
(hence on average once in every 25 runs) a near-unit entanglement.
The average harvested entanglement per run is 
$\bar{E}_v \equiv \sum_i p_i {\EF}_{v,i} = 0.07$.

It is especially noteworthy that in the case where $n_2=1$, the outer
modes are highly entangled, and the harvested qubits close to a Bell state.
This is owing to the properties of $\ket{\sigma_1}$. 
In the case where the central mode is unobserved and a trace taken,
the situation is comparable to that of a source which sends out pairs
of qubits, such that each pair is in a Bell state, but the receivers do not
know which, and in consequence they have a mixed state with zero entanglement.
But if they are informed which Bell state was furnished, they can recover
the entanglement. 

\begin{figure}

\begin{tikzpicture}[scale=0.95] \input{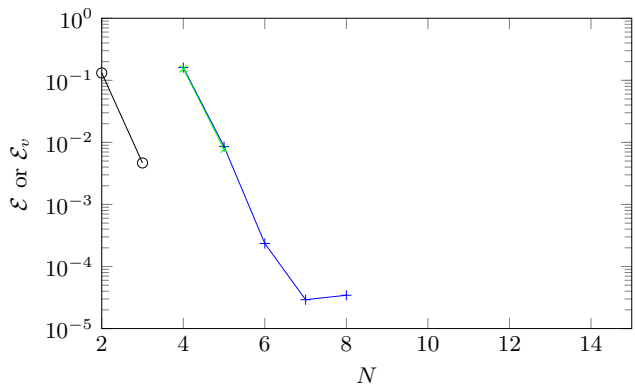} \end{tikzpicture}

\caption{Entanglement of the outer modes in a length-$N$ harmonic chain,
	after tracing over the inner modes.
black $\circ$: $\EF$ of outermost modes $[1;N]$ after tracing over $2\cdots (N\!-\!1)$ (this is zero for $N>3$);
blue $+$: upper bound on $\EF_v$ for outermost mode pairs $[\{1,2\};\{N\!-\!1,N\}]$ after tracing over $3\cdots (N\!-\!2)$;
($\EF_v$ is close to this upper bound for $N \le 5$.)}
\label{f.entbare}
\end{figure}

\begin{figure}
	
\begin{tikzpicture}[scale=0.95] \input{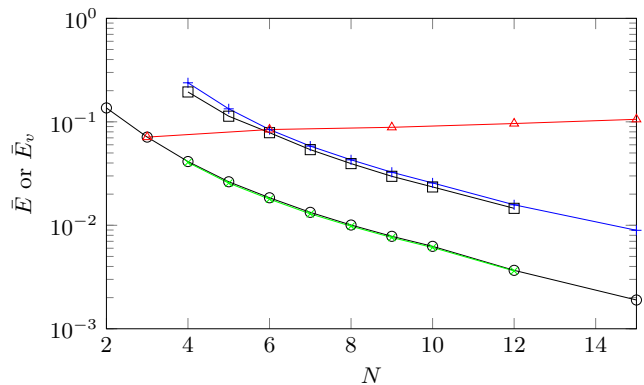} \end{tikzpicture}

\caption{Unlocked (heralded) entanglement.
$\circ$: average heralded entanglement $\bar{E}$ of modes ($1;N$);
blue $+$: average heralded entanglement of mode pairs;
green $\times$, black $\square$: harvested values;
red $\triangle$: $\bar{E}$ for a three-part partition (see text).
The values depend somewhat on the local $\omega_i$;
these were adjusted to maximise $\bar{E}$.}
\label{f.ent}
\end{figure}

Figs. \ref{f.entbare} and \ref{f.ent} extend the results to the range
$N = 4$--$15$ (limited by computational resources). We consider 
bipartite entanglement in three cases:
(i) between the two outer modes ($1$ and $N$); (ii) between 
the two outer pairs of modes
(i.e. modes $\{1,2\}$ and $\{N-1,N\}$); (iii) an
equal partition of the modes into three groups of equal size (when $N$ is a multiple of $3$).
In each case the remaining modes are either traced-over (Fig. \ref{f.entbare})
or measured and the result used as herald (Fig. \ref{f.ent}).
Fig. \ref{f.entbare} shows that in the absence
of information from the central modes, the entanglement is very small or zero.
Fig. \ref{f.ent} shows that the herald makes available a greatly enhanced entanglement.

We show results for a herald which can determine the number state (Fock state) 
of each of the central modes. In practice it is largely sufficient for the 
herald to distinguish $n=0$ from $n=1$,
so a harvesting operation would suffice to enable the required measurement. 
Also, one does not necessarily require
the $n$ values of all the inner modes; it largely suffices to determine whether
they sum to an even or an odd number.

The distribution of values of $\EF_i$ consists of a few values of order 1 and a large number of small values. Therefore the curves in Fig. \ref{f.ent} also indicate, approximately, the probability of obtaining $\EF \sim 1$ on any single run. 

The results for the third case (where the modes are
divided into three groups of $N/3$ each, with the central group acting as
herald) are shown by the red triangles in Fig. \ref{f.ent}. 
We find that for this case the heralded entanglement increases slowly with $N$,
for values of $N$ up to 15. This suggests this entanglement will not
fall at larger $N$; if so then we have a simple
way to obtain more than $0.1$ ebits at a distance $N/3$ on a longer chain.

It is important to note that the actions of the heralds and detectors
consist purely of local operations and classical communication.
Therefore they do not themselves cause the entanglement, but rather permit
entanglement present in the ground state to be accessed. Also, although
the outer observers must be timelike-separated from the inner ones in order
to receive their signals, they can remain spacelike-separated from one another. 
Furthermore, when the classical signals arrive, the qubits acting as detectors 
can already have been isolated from the set of oscillators, again
ruling out the suspicion
that the entanglement was communicated along the chain.

So far we considered the case where the central local modes are either
all measured, or all traced-over. We next consider the case where some but
not all are measured. This may be important to the extension to the continuous
case. We investigated the heralded entanglement between mode-pairs $\{1,2\}$
and $\{N-1,N\}$ when modes $3$ and $N-2$ are traced-over, and the
remaining $N-6$ (i.e. modes $4$ to $N-3$) act as herald. 
At $N=6$ in this case there is no herald, and the result
$\EF_v = 2.4 \times 10^{-4}$ forms a data point on  
Fig. \ref{f.entbare}. For $N$ in the range $7$ to $10$ our numerical method
was not able to give precise values, but indicated that the entanglement
falls exponentially from the $N=6$ value at a decay rate similar to
that shown in Fig. \ref{f.ent}, yielding values of order $10^{-4}$. Though
small, this is very much larger (perhaps infinitely larger) than the
unheralded value.

An experiment to demonstrate the ideas in this paper can be performed with
existing technology in a linear ion trap, largely as described in \cite{Retzker2005}. For trapped ions one replaces (\ref{w1}) by
$\omega^2_1=\omega^2_3= \nu_2(3\nu_1\nu_3+(2\nu_1 + \nu_3)\omega_2)/
(\nu_1 + 2 \nu_3 + 3 \omega_2)$.
The concepts also apply in principle to any system with normal modes of vibration, such as molecules and crystals.
Since a harmonic chain
can serve as a discretized model of a bosonic quantum field, we expect
the same qualitative behaviour for the vacuum state of such a field. That
this is so ought to be implied by existing theoretical treatments but the authors have
not found it clearly pointed out in the literature. In any case equations 
(\ref{sig0}), (\ref{sig1}) and the general approach adopted here are new.

Many of the vibrational modes of molecules 
have negligible thermal excitation at room temperature.
The following research questions are suggested. 
(i) Extend Fig. \ref{f.ent} to higher $N$. (ii) A quantitative analysis
of the continuous (i.e. quantum field) limit. (iii) Are there fast processes,
such as chemical reactions, that could harvest vibrational entanglement of a molecule into some other degree of freedom, such as electron spin? (iv) Does such entanglement exist in molecules dissolved or floating in a liquid such as water at STP? (v) If so, then can it be harvested, for example by a reaction whose timescale is fast compared to the vibrational relaxation time in a liquid
environment?

\bibliography{myrefs}

@book{BkNielsen,
  author = {Michael A. Nielsen and Isaac L. Chuang},
  title = "Quantum Computation and Quantum Information",
  publisher = {Cambridge University Press},
  address = {Cambridge},
  year = {2000}
}

@article{Unruh1976,
  title = {Notes on black-hole evaporation},
  author = {Unruh, W. G.},
  journal = {Phys. Rev. D},
  volume = {14},
  issue = {4},
  pages = {870--892},
  numpages = {0},
  year = {1976},
  month = {Aug},
  publisher = {American Physical Society},
  doi = {10.1103/PhysRevD.14.870},
  url = {https://link.aps.org/doi/10.1103/PhysRevD.14.870}
}

@Article{Reznik2003,
author={Reznik, Benni},
title={Entanglement from the Vacuum},
journal={Foundations of Physics},
year={2003},
month={Jan},
day={01},
volume={33},
number={1},
pages={167-176},
issn={1572-9516},
doi={10.1023/A:1022875910744},
url={https://doi.org/10.1023/A:1022875910744}
}

@article{Reznik2005,
  title = {Violating Bell's inequalities in vacuum},
  author = {Reznik, Benni and Retzker, Alex and Silman, Jonathan},
  journal = {Phys. Rev. A},
  volume = {71},
  issue = {4},
  pages = {042104},
  numpages = {4},
  year = {2005},
  month = {Apr},
  publisher = {American Physical Society},
  doi = {10.1103/PhysRevA.71.042104},
  url = {https://link.aps.org/doi/10.1103/PhysRevA.71.042104}
}

@article{Martin-Martinez2016,
doi = {10.1088/1367-2630/18/4/043031},
url = {https://doi.org/10.1088/1367-2630/18/4/043031},
year = {2016},
month = {apr},
publisher = {IOP Publishing},
volume = {18},
number = {4},
pages = {043031},
author = {Martín-Martínez, Eduardo and Sanders, Barry C},
title = {Precise space–time positioning for entanglement harvesting},
journal = {New Journal of Physics}
}

@article{Gooding2024,
doi = {10.1088/1367-2630/ad8675},
url = {https://doi.org/10.1088/1367-2630/ad8675},
year = {2024},
month = {oct},
publisher = {IOP Publishing},
volume = {26},
number = {10},
pages = {105001},
author = {Gooding, Cisco and Sachs, Allison and Mann, Robert B and Weinfurtner, Silke},
title = {Vacuum entanglement probes for ultra-cold atom systems},
journal = {New Journal of Physics}
}

@article{Lindel2024,
  title = {Entanglement harvesting from electromagnetic quantum fields},
  author = {Lindel, Frieder and Herter, Alexa and Gebhart, Valentin and Faist, J\'er\^ome and Buhmann, Stefan Yoshi},
  journal = {Phys. Rev. A},
  volume = {110},
  issue = {2},
  pages = {022414},
  numpages = {19},
  year = {2024},
  month = {Aug},
  publisher = {American Physical Society},
  doi = {10.1103/PhysRevA.110.022414},
  url = {https://link.aps.org/doi/10.1103/PhysRevA.110.022414}
}

@article{Gao2025,
  title = {Detecting spacelike vacuum entanglement at all distances and promoting negativity to a necessary and sufficient entanglement measure in many-body regimes},
  author = {Gao, Boyu and Klco, Natalie},
  journal = {Phys. Rev. A},
  volume = {112},
  issue = {1},
  pages = {012430},
  numpages = {13},
  year = {2025},
  month = {Jul},
  publisher = {American Physical Society},
  doi = {10.1103/m9w1-ppqz},
  url = {https://link.aps.org/doi/10.1103/m9w1-ppqz}
}

@article{Audenaert2002,
  title = {Entanglement properties of the harmonic chain},
  author = {Audenaert, K. and Eisert, J. and Plenio, M. B. and Werner, R. F.},
  journal = {Phys. Rev. A},
  volume = {66},
  issue = {4},
  pages = {042327},
  numpages = {14},
  year = {2002},
  month = {Oct},
  publisher = {American Physical Society},
  doi = {10.1103/PhysRevA.66.042327},
  url = {https://link.aps.org/doi/10.1103/PhysRevA.66.042327}
}

@article{Botero2003,
  title = {Modewise entanglement of Gaussian states},
  author = {Botero, Alonso and Reznik, Benni},
  journal = {Phys. Rev. A},
  volume = {67},
  issue = {5},
  pages = {052311},
  numpages = {5},
  year = {2003},
  month = {May},
  publisher = {American Physical Society},
  doi = {10.1103/PhysRevA.67.052311},
  url = {https://link.aps.org/doi/10.1103/PhysRevA.67.052311}
}

@article{Retzker2005,
  title = {Detecting Vacuum Entanglement in a Linear Ion Trap},
  author = {Retzker, A. and Cirac, J. I. and Reznik, B.},
  journal = {Phys. Rev. Lett.},
  volume = {94},
  issue = {5},
  pages = {050504},
  numpages = {4},
  year = {2005},
  month = {Feb},
  publisher = {American Physical Society},
  doi = {10.1103/PhysRevLett.94.050504},
  url = {https://link.aps.org/doi/10.1103/PhysRevLett.94.050504}
}

@article{Caves1991,
  title = {Photon statistics of two-mode squeezed states and interference in four-dimensional phase space},
  author = {Caves, Carlton M. and Zhu, Chang and Milburn, G. J. and Schleich, W.},
  journal = {Phys. Rev. A},
  volume = {43},
  issue = {7},
  pages = {3854--3861},
  numpages = {0},
  year = {1991},
  month = {Apr},
  publisher = {American Physical Society},
  doi = {10.1103/PhysRevA.43.3854},
  url = {https://link.aps.org/doi/10.1103/PhysRevA.43.3854}
}

@article{Tserkis2017,
  title = {Quantifying entanglement in two-mode Gaussian states},
  author = {Tserkis, Spyros and Ralph, Timothy C.},
  journal = {Phys. Rev. A},
  volume = {96},
  issue = {6},
  pages = {062338},
  numpages = {6},
  year = {2017},
  month = {Dec},
  publisher = {American Physical Society},
  doi = {10.1103/PhysRevA.96.062338},
  url = {https://link.aps.org/doi/10.1103/PhysRevA.96.062338}
}

@article{Summers1987,
    author = {Summers, Stephen J. and Werner, Reinhard},
    title = {Bell's inequalities and quantum field theory. I. General setting},
    journal = {Journal of Mathematical Physics},
    volume = {28},
    number = {10},
    pages = {2440-2447},
    year = {1987},
    month = {10},
    issn = {0022-2488},
    doi = {10.1063/1.527733},
    url = {https://doi.org/10.1063/1.527733}
}

@article{Botero2004,
  title = {Spatial structures and localization of vacuum entanglement in the linear harmonic chain},
  author = {Botero, Alonso and Reznik, Benni},
  journal = {Phys. Rev. A},
  volume = {70},
  issue = {5},
  pages = {052329},
  numpages = {21},
  year = {2004},
  month = {Nov},
  publisher = {American Physical Society},
  doi = {10.1103/PhysRevA.70.052329},
  url = {https://link.aps.org/doi/10.1103/PhysRevA.70.052329}
}

@article{Milburn1984,
doi = {10.1088/0305-4470/17/4/015},
url = {https://doi.org/10.1088/0305-4470/17/4/015},
year = {1984},
month = {mar},
publisher = {},
volume = {17},
number = {4},
pages = {737},
author = {G J Milburn},
title = {Multimode minimum uncertainty squeezed states},
journal = {Journal of Physics A: Mathematical and General}
}

@article{Caves1985,
  title = {New formalism for two-photon quantum optics. I. Quadrature phases and squeezed states},
  author = {Caves, Carlton M. and Schumaker, Bonny L.},
  journal = {Phys. Rev. A},
  volume = {31},
  issue = {5},
  pages = {3068--3092},
  numpages = {0},
  year = {1985},
  month = {May},
  publisher = {American Physical Society},
  doi = {10.1103/PhysRevA.31.3068},
  url = {https://link.aps.org/doi/10.1103/PhysRevA.31.3068}
}

@article{Wootters1998,
  title = {Entanglement of Formation of an Arbitrary State of Two Qubits},
  author = {Wootters, William K.},
  journal = {Phys. Rev. Lett.},
  volume = {80},
  issue = {10},
  pages = {2245--2248},
  numpages = {0},
  year = {1998},
  month = {Mar},
  publisher = {American Physical Society},
  doi = {10.1103/PhysRevLett.80.2245},
  url = {https://link.aps.org/doi/10.1103/PhysRevLett.80.2245}
}

\end{document}